\documentclass[reprint, amsmath, amssymb, aps, prstab, floatfix]{revtex4-1}

\usepackage{graphicx}
\usepackage{dcolumn}
\usepackage{bm}

\begin{document}

\title{Analytic plasma wakefield limits for active plasma lenses}

\author{Carl A. Lindstr\o m}
\email{c.a.lindstrom@fys.uio.no}
\affiliation{Department of Physics, University of Oslo, 0316 Oslo, Norway}
\author{Erik Adli}
\affiliation{Department of Physics, University of Oslo, 0316 Oslo, Norway}

\date{\today}

\begin{abstract}
	Active plasma lensing is a promising technology for compact focusing of particle beams that has seen a recent surge of interest. While these lenses can provide strong focusing gradients of order kT/m and focusing in both transverse planes, there are limitations from nonlinear aberrations, causing emittance growth in the beams being focused. One cause of such aberrations is beam-driven plasma wakefields, present if the beam density is sufficiently high. We develop simple, but powerful analytic formulas for the effective focusing gradient from these wakefields, and use this to set limits on which parts of the beam and plasma parameter space permits distortion-free use of active plasma lenses. It is concluded that the application of active plasma lenses to conventional and plasma-based linear colliders may prove very challenging, except perhaps in the final focus system, unless the typical discharge currents used are dramatically increased, and that in general these lenses are better suited for accelerator applications with lower beam intensities.
\end{abstract}

\maketitle

\section{Introduction}

Active plasma lenses \cite{vanTilborgPRL2015} have seen a recent rise in interest, even though the first prototype was built by W. Panofsky and Baker already in 1950 \cite{PanofskyRSI1950}. This surge is mainly linked to novel accelerator concepts, which require strong and compact focusing beyond what conventional quadrupoles usually provide. A key difference is that quadrupoles focus in one plane while defocusing in the other, whereas active plasma lenses focus (or defocus) in both planes simultaneously. Such azimuthally symmetric focusing allows the focal length to be drastically shortened, which is important e.g.~in capturing the highly divergent beams produced by plasma accelerators.

However, azimuthally symmetric focusing requires circular magnetic field lines around the longitudinal axis, which can only exist in the presence of a longitudinal current density. To conduct this current density, the active plasma lens uses a tenuous plasma to also allow a particle beam to pass through. Following this, several problems emerge: beam scattering from collisions with ions, uneven heating of the plasma giving transversely nonuniform conductivity and therefore nonlinear focusing forces \cite{vanTilborgPRAB2017}, as well as plasma wakefields excited by the beam setting up potentially very strong, nonlinear focusing fields. In this paper, we develop analytic limits for the latter problem, aimed as a guide for designing an appropriate active plasma lens given a set of beam parameters such that the distortion from plasma wakefields is negligible. We therefore avoid considering the detailed beam optics needed to calculate emittance growth: if there is no significant nonlinear aberration, there is also negligible emittance growth.

\section{Active plasma lensing}

Active plasma lenses are current-based magnetic lenses with focusing fields of the order kT/m. They are called \textit{active} to distinguished them from \textit{passive} plasma lenses, which require no actively driven external current, but instead rely on the plasma wakefields driven by the beam itself. Although passive plasma lenses can provide much stronger focusing fields of order MT/m, the fields are generally not transversely or longitudinally uniform (unless driven by a leading laser or particle beam) and is only focusing for negatively charged beams. An ideal active plasma lens, however, does provide a transversely and longitudinally uniform focusing field, and can focus both negatively and positively charged beams.

Consider a cylindrical capillary filled with a conducting medium (a plasma) of uniform conductivity. An external current $I$ between electrodes on the entry and exit of this capillary will result in a uniform current density $J_z = I/(\pi R^2)$, where $R$ is the radius of the capillary. Assuming longitudinal and azimuthal symmetry, Ampere's law reduces to
\begin{equation}
	\label{eq:AmpLaw}
    \frac{1}{r} \frac{\partial ( r B_\phi )}{\partial r} = \mu_0 J_z,
\end{equation}
where $B_\phi$ is the azimuthal magnetic field and $r$ is the radial distance from the axis. Multiply Eq.~\ref{eq:AmpLaw} by $r$ and integrate, assuming that $J_z$ is independent of $r$, to give
\begin{equation}
    B_\phi = \frac{\mu_0 J_z r}{2}.
\end{equation}
Differentiate and substitute for the current density to end up with an expression for the magnetic field gradient in the active plasma lens (APL)
\begin{equation}
    g_{\mathrm{APL}} = \frac{\partial B_\phi}{\partial r} = \frac{\mu_0 I}{2 \pi R^2}.
\end{equation}

This relation is correct only for perfectly uniform current densities. The assumption of uniform current breaks down if the plasma density or temperature varies radially, typically caused by heating with thermally conductive capillary walls or by incomplete ionization. Reference \cite{vanTilborgPRAB2017} reports that this nonuniformity is present in state-of-the-art active plasma lens designs, and that both theory and experiment points to an enhancement of the field strength on axis of a small factor. However, since this is still within the same order of magnitude and independent of plasma wakefields, we will ignore it in further considerations.

\section{Plasma wakefields}

Beam-plasma interactions are a complicated matter. However, our goal is simple: to determine when the focusing strength of unwanted plasma wakefields are no longer negligible compared to that of the active plasma lens. In the end, this will manifest in inequalities setting limits on combinations of beam and plasma lens parameters. 

The premise of the plasma wakefield accelerator concept \cite{ChenPRL1985,RuthPA1985,JoshiPT2003} is that a plasma can support very strong electromagnetic fields. Very intense, high charge density bunches are required to probe the nonlinear limit of these fields, in what is known as the \textit{blowout} regime \cite{RosenzweigPRA1991}. In this regime, the focusing wakefield may be orders of magnitude stronger than that of the active plasma lens. This means that for most reasonable beam parameters where an active plasma lens can be used, the beam density must be low and we are instead in the perturbative \textit{linear} regime, which can be treated analytically.

\subsection{Perturbative linear regime}
The linear theory is a straightforward perturbative model, which can be described in varying levels of detail. We require at least a description of transverse focusing fields, and for simplicity we will work in cylindrically symmetric geometry. Reference \cite{BlumenfeldThesis2009} outlines exactly such a 2D linear theory, and we will simply import its conclusions. The cylindrical geometry means we are not able to explicitly describe flat beams (important e.g. in collider applications), but we will later show that the expressions are useful also in such cases.

Consider a Gaussian beam of $N$ particles and root mean square (rms) bunch length $\sigma_z$ and rms transverse size $\sigma_r$ in both transverse planes, with a particle density
\begin{equation}
	\label{eq:GaussBeamDens}
	n_b(z,r) = -\frac{N e^{-\frac{r^2}{2\sigma_r^2} - \frac{z^2}{2\sigma_z^2}}}{\sqrt{2\pi}^3 \sigma_r^2\sigma_z},
\end{equation}
using a minus sign to signify use of electrons. This will cause a plasma density perturbation
\begin{equation}
	\label{eq:PlasmaPerturb}
	\delta n(z,r) = - k_p \int_z^\infty n_b(z',r) \sin k_p (z - z') dz',
\end{equation}
which is simply the convolution of the plasma Green's function of all the beam charge in front. The characteristic plasma wavenumber is given by
\begin{equation}
	k_p = \sqrt{\frac{n_0 e^2}{\epsilon_0 m_e c^2}},
\end{equation}
where $e$ and $m_e$ is the electron charge and mass, $\epsilon_0$ and $c$ is the vacuum permittivity and light speed, and $n_0$ is the plasma density. Immediately, we see that a Gaussian beam does not give a closed-form expression for $\delta n$, but involves an integral. To simplify, we will therefore consider two separate regimes: long beams, where $k_p \sigma_z \gg 1$, and short beams, where $k_p \sigma_z \ll 1$. In the transition region where $k_p \sigma_z \approx 1$ only a rough estimate is provided.

For each of these two regimes, the resulting plasma density perturbation is entered in the linear theory to give an expression for the transverse electric field
\begin{equation}
	\label{eq:TransField}
	E_r(z,r) = \frac{e}{\epsilon_0} \int_0^\infty \frac{\partial \delta n(z,r')}{\partial r} K_1 (k_p r_>) I_1 (k_p r_<) r' dr',
\end{equation}
where $I_n$ and $K_n$ are the $n$th order modified Bessel functions of the first and second kind, respectively, and $r_>$ is the greater of $r$ and $r'$, and $r_<$ is the lesser. We are however mostly interested in the transverse gradient
\begin{equation}
	\label{eq:TransGradient}
	g_r(z,r) = \frac{1}{c} \frac{\partial E_r(z,r)}{\partial r}.
\end{equation}
Note that we have normalized the gradient by $c$ to be comparable to the magnetic field gradient in the active plasma lens (measured in units of T/m).

In a Gaussian beam, the maximum focusing strength will always occur on axis, $r = 0$, as this is where the beam is densest and therefore the plasma perturbation largest. This will simplify our calculations greatly, as we only need to work with $g_r(z,0)$.

\begin{figure*}[t]
	\centering\includegraphics[width=\textwidth]{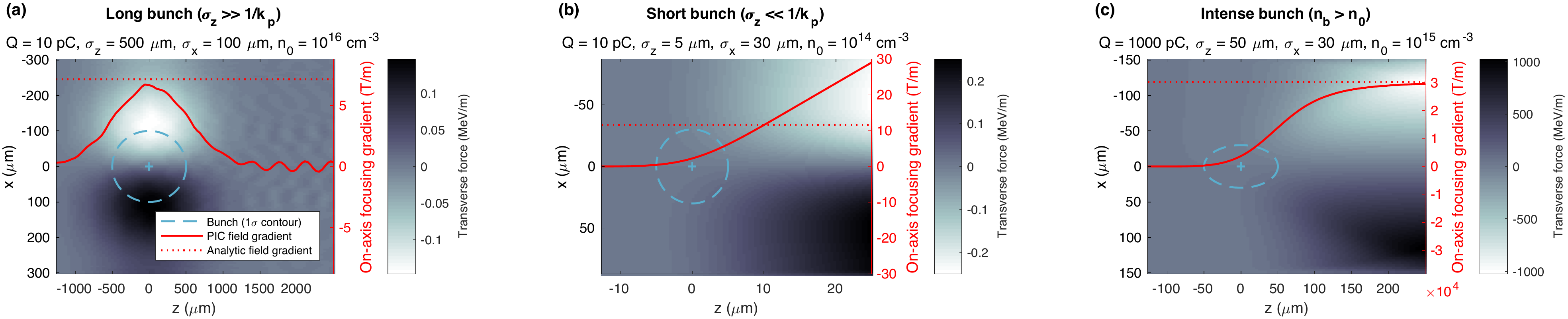}
	\caption{Particle-in-cell (PIC) simulations of bunches in the three regimes considered: long (a), short (b) and intense (c) bunches. Each simulation shows a density plot of the transverse focusing force and an indication of the electron bunch location with a 1$\sigma$ contour (dashed blue line). Overlaid is the PIC on-axis field gradient (solid red line) measured in units of magnetic field gradient, as well as the analytic estimate (Eqs.~\ref{eq:LongBunchMaxField}, \ref{eq:ShortBunchMaxField} and \ref{eq:IntenseBunchMaxField} respectively) of the maximum gradient (dashed red line). Simulations were run using QuickPIC Open Source \cite{AnJCP2013}.}
    \label{fig:SimRegimes}
\end{figure*}

\subsubsection{Long bunches}
We start with long beams, as they are conceptually simpler. In the case when $\sigma_z \gg 1/k_p$, the plasma has sufficient time to restore overall neutrality, and hence $\delta n = -n_b$. Equation \ref{eq:TransField} requires a transverse derivative:
\begin{equation}
	\frac{\partial \delta n}{\partial r} = -\frac{\partial n_b}{\partial r} = \frac{N e^{-\frac{z^2}{2\sigma_z^2}}}{\sqrt{2\pi}^3 \sigma_r^2\sigma_z} \left( -\frac{r}{\sigma_r^2} \right) e^{- \frac{r^2}{2\sigma_r^2}}.
\end{equation}
We enter this derivative into Eq.~\ref{eq:TransGradient} and find
\begin{equation}
	\label{eq:GradientLongIncomplete}
	g_r(z,r) = A(z) \frac{\partial}{\partial r} \int_0^\infty {r'}^2 e^{- \frac{{r'}^2}{2\sigma_r^2}} K_1 (k_p r_>) I_1 (k_p r_<) r' dr',
\end{equation}
where $A(z)$ is a $z$-dependent factor given by
\begin{equation}
	A(z) = -\frac{e N \mu_0 c e^{-\frac{z^2}{2\sigma_z^2}}}{\sqrt{2\pi}^3 \sigma_r^4 \sigma_z}
\end{equation}
and we used the identity $\mu_0 \epsilon_0 c^2 = 1$. Equation~\ref{eq:GradientLongIncomplete} must be split into integrals with limits above and below $r$
\begin{equation}\begin{split}
	\label{eq:TransGradSplit}
	g_r(z,r) = A(z) \frac{\partial}{\partial r} \int_0^r {r'}^2 e^{- \frac{{r'}^2}{2\sigma_r^2}} K_1(k_p r) I_1(k_p r') dr' \\ + A(z) \frac{\partial}{\partial r} \int_r^\infty {r'}^2 e^{- \frac{{r'}^2}{2\sigma_r^2}} K_1(k_p r') I_1(k_p r) dr',
\end{split}\end{equation}
We apply Leibniz rule for differentiation under the integral sign to the two terms, giving
\begin{equation}\begin{split}
	\label{eq:BessIntLower}
	\frac{\partial}{\partial r} \int_0^r {r'}^2 e^{- \frac{{r'}^2}{2\sigma_r^2}} K_1(k_p r) I_1(k_p r') dr' = \\ \frac{k_p}{2}(K_0(k_p r)+K_2(k_p r)) \int_0^r {r'}^2 e^{- \frac{{r'}^2}{2\sigma_r^2}} I_1(k_p r') dr' \\ + {r}^2 e^{- \frac{r^2}{2\sigma_r^2}} K_1(k_p r) I_1(k_p r), 
\end{split}\end{equation}
and similarly
\begin{equation}\begin{split}
	\label{eq:BessIntUpper}
	\frac{\partial}{\partial r} \int_r^\infty {r'}^2 e^{- \frac{{r'}^2}{2\sigma_r^2}} K_1(k_p r') I_1(k_p r) dr' = \\ \frac{k_p}{2}(I_0(k_p r)+I_2(k_p r)) \int_r^\infty {r'}^2 e^{- \frac{{r'}^2}{2\sigma_r^2}} K_1(k_p r') dr' \\ - {r}^2 e^{- \frac{r^2}{2\sigma_r^2}} K_1(k_p r) I_1(k_p r),
\end{split}\end{equation}
where we have used the Bessel identities $\frac{\partial}{\partial x} K_n(x) = -\frac{1}{2}(K_{n-1}(x) + K_{n+1}(x))$ and $\frac{\partial}{\partial x} I_n(x) = \frac{1}{2}(I_{n-1}(x) + I_{n+1}(x))$.

The maximum of this gradient $g_r$ is along the longitudinal axis of the beam ($r = 0$). We therefore evaluate the central field gradient $g_r(z,0)$, which cancels all of Eq.~\ref{eq:BessIntLower} and the last term in Eq.~\ref{eq:BessIntUpper} to give
\begin{equation}
	\label{eq:CentralTransField}
	g_r(z,0) = A(z) \frac{k_p}{2} \int_0^\infty {r'}^2 e^{- \frac{{r'}^2}{2\sigma_r^2}} K_1(k_p r') dr',
\end{equation}
where we have used $I_0(0) = 1$ and $I_2(0) = 0$. Additionally, as seen in Fig.~\ref{fig:SimRegimes}(a), the maximum in the longitudinal is in the center of the beam ($z=0$): the maximum field is therefore $g_r(0,0)$. This is a known integral which evaluates to
\begin{equation}
	\label{eq:LongBuchGIncomplete}
	g_r(0,0) = A(0) \frac{{\sigma_r}^2}{2} \left(  1 + \frac{k_p^2 \sigma_r^2}{2} e^{\frac{k_p^2 \sigma_r^2}{2}} \mathrm{Ei}\left( -\frac{k_p^2 \sigma_r^2}{2} \right) \right),
\end{equation}
where we have used the standard \textit{exponential integral}
\begin{equation}
	\mathrm{Ei}(x) = -\int_{-x}^\infty \frac{e^{-t}}{t} dt.
\end{equation}
Equation \ref{eq:LongBuchGIncomplete} can be simplified by defining a new function
\begin{equation}
	\label{eq:ChiSubFunction}
	\chi(x) = 1 + \frac{x^2}{2} e^{\frac{x^2}{2}} \mathrm{Ei}\left( -\frac{x^2}{2} \right).
\end{equation}
Finally we expand $A(0)$ to give the maximum focusing field gradient for long beams
\begin{equation}
	\label{eq:LongBunchMaxField}
	g^{\mathrm{long}}_{\mathrm{max}} = -\frac{e N c \mu_0}{2 \sqrt{2\pi}^3 \sigma_r^2 \sigma_z} \chi(k_p\sigma_r).
\end{equation}

\begin{figure*}[ht]
	\centering\includegraphics[width=0.9\textwidth]{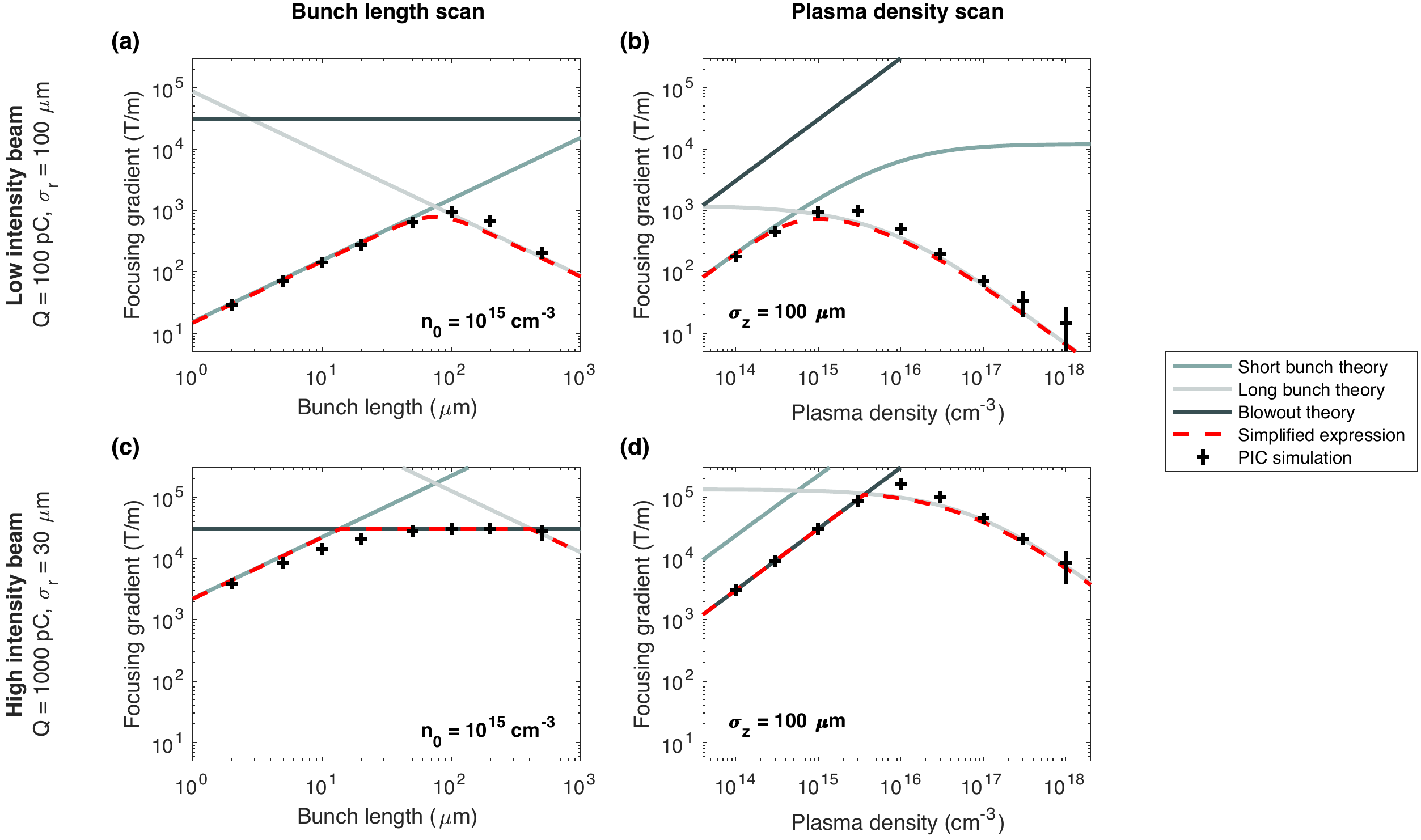}
	\caption{Parameter scans of bunch length (a and c) and plasma density (b and d), showing that the expressions agree with particle-in-cell (PIC) simulations (black crosses) regarding the maximum focusing gradient within the bunch. This works both for low intensity beams (a and b), which are in the linear regime (turquoise and gray lines), and for high intensity beams (c and d), which are also in the nonlinear blowout regime (black lines). The simplified expression (Eq.~\ref{eq:MaxFieldSimplified}) is also shown (red dashed line). Error bars quantify the rms of the noise in the simulation, which can occur when the beam spans a large number of plasma wavelengths.}
    \label{fig:Scans}
\end{figure*}

\subsubsection{Short bunches}
Short bunches do not allow the plasma to reach equilibrium before the bunch has passed, and we therefore have to use Eq.~\ref{eq:PlasmaPerturb} to find the density perturbation. Since the beam is short, all longitudinal distances are shorter than $k_p^{-1}$, which allows us to make use of the small angle approximation to remove the sinusoidal dependence
\begin{equation}
	\delta n(z,r) \approx \frac{N e^{-\frac{r^2}{2\sigma_r^2}} k_p^2}{\sqrt{2\pi}^3 \sigma_r^2\sigma_z} \int_z^\infty e^{-\frac{{z'}^2}{2\sigma_z^2}} (z - z') dz'.
\end{equation}
This integral increases monotonically going backwards through the bunch, quickly approximating a linear increase
\begin{equation}
	\label{eq:MonotonicToLinear}
	\lim_{z \rightarrow -\infty} \int_z^\infty e^{-\frac{{z'}^2}{2\sigma_z^2}} (z - z') dz' = \sqrt{2 \pi} z,
\end{equation}
which means that we must simply define a location to be the back of the bunch, say at $z = -2 \sigma_z$, for which the approximation in Eq.~\ref{eq:MonotonicToLinear} is very good (see Fig.~\ref{fig:SimRegimes}(b)). This gives a maximum density perturbation
\begin{equation}
	\delta n(-2\sigma_z,r) = -\frac{N e^{-\frac{r^2}{2\sigma_r^2}} k_p^2 \sigma_z }{\pi \sigma_r^2}.
\end{equation}
We observe that this density perturbation relates to the density perturbation $-n_b$ for long beams via 
\begin{equation}
	\label{eq:LongShortRatio}
	\delta n(-2\sigma_z,r) = - n_b(z,r) \sqrt{8 \pi} k_p^2 \sigma_z^2 e^{\frac{z^2}{2\sigma_z^2}}.
\end{equation}
Note that the short bunch density perturbation does not have any additional $r$-dependence, implying that we can reuse the conclusion from the long bunch calculation, by simply multiplying the factor $A(z)$ by the ratio Eq.~\ref{eq:LongShortRatio} to give
\begin{equation}
	A(-2\sigma_z) = -\frac{e N \mu_0 c k_p^2 \sigma_z}{\pi \sigma_r^4}.
\end{equation}
Substituting this into Eq.~\ref{eq:CentralTransField} finally gives an expression for the maximum focusing field affecting a short bunch
\begin{equation}
	\label{eq:ShortBunchMaxField}
	g^{\mathrm{short}}_{\mathrm{max}} = -\frac{e N c \mu_0 k_p^2 \sigma_z}{2 \pi \sigma_r^2 }  \chi(k_p\sigma_r),
\end{equation}
where we have again employed the function $\chi(x)$. Importantly, while Eqs.~\ref{eq:LongBunchMaxField} and \ref{eq:ShortBunchMaxField} look similar, they scale differently with plasma density and bunch length.

\subsection{Nonlinear blowout regime}
A semi-analytic description of the blowout regime exists \cite{LuPRL2006}, but is unnecessary for our purposes. For beams of very high density $n_b > n_0$, the plasma perturbation saturates (see Fig.~\ref{fig:SimRegimes}c), leaving only ions on axis. The focusing field is now transversely linear and the maximum gradient is given only by the exposed charge of the ion column
\begin{equation}
	\label{eq:IntenseBunchMaxField}
	g^{\mathrm{intense}}_{\mathrm{max}} = -\frac{e n_0}{2 c \epsilon_0}.
\end{equation}
This gradient is typically very strong in comparison with the active plasma lens gradient, and is the principle behind the passive plasma lens. However for a single bunch, the focusing is not uniform along the length of the bunch, since it takes the blowout some time to form, leaving the front of the bunch unfocused (also known as head erosion).

The combination of Eqs.~\ref{eq:LongBunchMaxField}, \ref{eq:ShortBunchMaxField} and \ref{eq:IntenseBunchMaxField} gives the maximum wakefield focusing for any charge, beam size, bunch length and plasma density:
\begin{equation}
	\label{eq:TotalBunchMaxField}
	g_{\mathrm{max}} = \min(g^{\mathrm{long}}_{\mathrm{max}}, g^{\mathrm{short}}_{\mathrm{max}}, g^{\mathrm{intense}}_{\mathrm{max}}).
\end{equation}

\subsection{Particle-in-cell benchmarking}

For verification, we compare these expressions to particle-in-cell (PIC) simulations. In this case, a single-step (non-evolving) calculation is sufficient, so it is suitable to use the fast, quasi-static code QuickPIC Open Source \cite{AnJCP2013}. The resolution and grid size is set to ensure at least 25 cells per plasma wavelength in each direction, rounded up to the closest power of 2 (and minimum 64 cells).

Figure~\ref{fig:Scans} shows a benchmarking of a bunch length scan and a plasma density scan for both low intensity (a and b) and high intensity beams (c and d). The expressions found are consistent with PIC simulations, to at worst approximately a factor 2, even over many orders of magnitude of parameter variation.

\begin{figure}[t]
	\centering\includegraphics[width=0.95\linewidth]{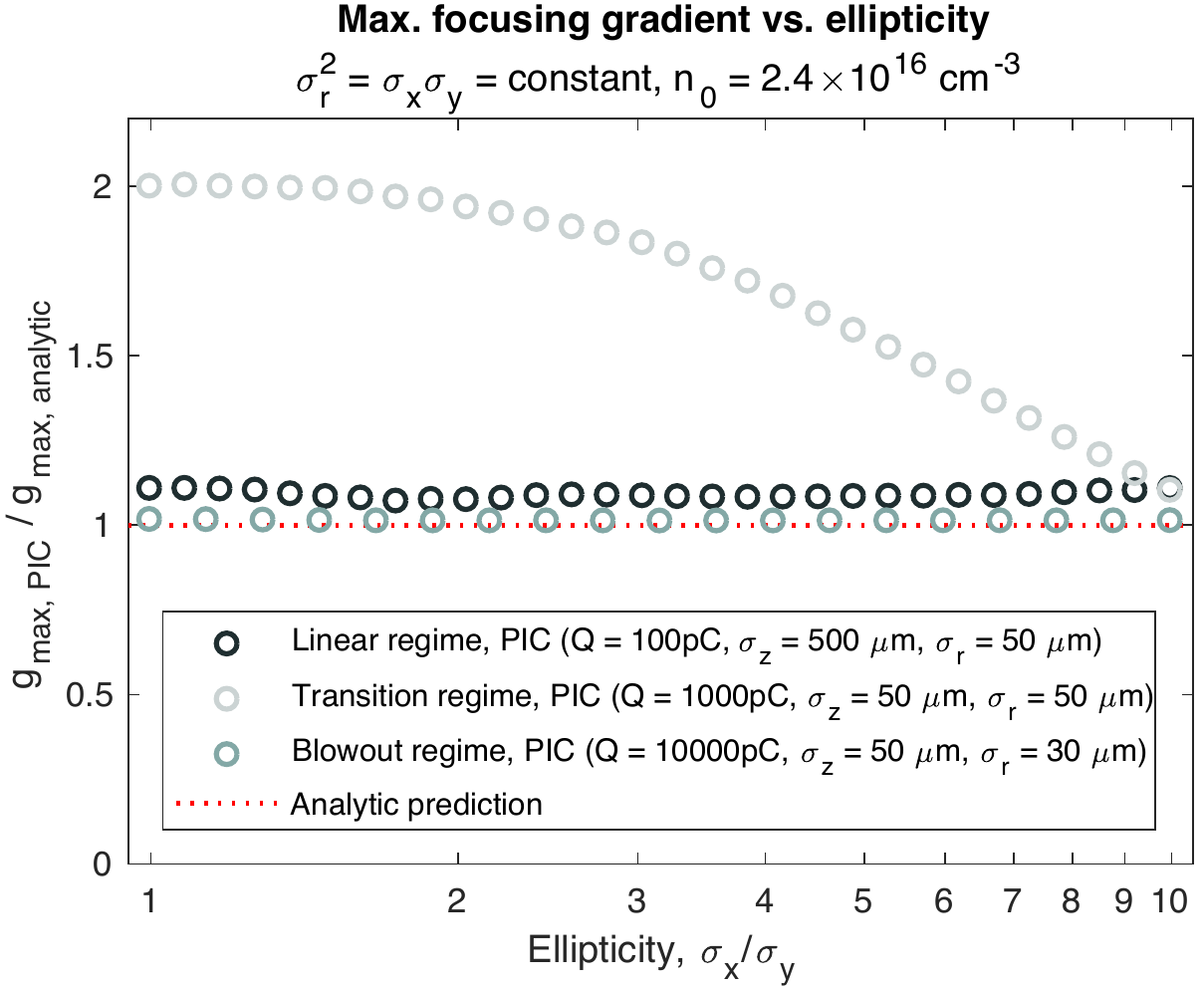}
	\caption{Scan of ellipticity (ratio of horizontal to vertical beam size) with a constant equivalent round beam size (Eq.~\ref{eq:EquivBeamSize}) using PIC simulations. In the linear (black circles) and the blowout regime (turquoise circles) the maximum plasma wakefield focusing gradient is independent of the ellipticity. However, in the transition region (gray circles), this independence only holds for small ellipticities, but not in general. Note that in this transition region the analytic prediction may deviate by a small factor: here by a factor 2.}
    \label{fig:Ellipticity}
\end{figure}

\subsection{Elliptical beams}

All the above considerations have assumed round beams and may therefore be of questionable applicability to elliptical beams of different beam sizes in the two transverse planes, as used e.g.~in a linear collider. In the linear regime, the maximum focusing field gradient is a quantity mainly dependent on the local beam density at the center of the bunch. We can therefore calculate the equivalent beam size as if the beam was round using the geometric mean of the beam sizes in the two transverse planes
\begin{equation}
	\label{eq:EquivBeamSize}
	\sigma_r = \sqrt{\sigma_x \sigma_y}.
\end{equation}
The independence also holds for the blowout regime, as the focusing gradient is only determined by the plasma density. However, in the transition between these two extremes, elliptical beams excite a nonstandard plasma perturbation, and we therefore do not expect the round-beam prediction to remain valid. These conclusions are supported by Figure \ref{fig:Ellipticity}, which shows a comparison between PIC-simulated elliptical beams and the analytic prediction for a constant $\sigma_r$.

To avoid distortions due to plasma wakefields, we require low density beams (bar a few exceptions including ultra-short beams). This implies that the linear regime is the most relevant regime for estimating limits, and we can therefore safely consider flat collider beams.

\section{Simplified analytic expressions}

\subsection{Maximum plasma wakefield focusing gradient}

Our overall goal is to find an analytic and fast-to-calculate expression for the plasma wakefield-based focusing gradient. Equation \ref{eq:TotalBunchMaxField} is already many orders of magnitude faster to calculate than a PIC simulation, but is still not a closed-form expression. This is because it contains the exponential integral $\mathrm{Ei}(x)$, which must be evaluated numerically. Trading accuracy for speed of calculation, we can approximate the substitution function $\chi(x)$ (see Eq.~\ref{eq:ChiSubFunction}) by considering its asymptotic limits
\begin{eqnarray}
	\lim_{x \to 0} \chi(x) &=& 1, \\
    \lim_{x \to \infty} \chi(x) &=& \frac{2}{x^2}.
\end{eqnarray}
These can be combined into 
\begin{equation}
	\label{eq:ChiApprox}
	\chi(x) \approx \left( 1 + \frac{x^2}{2} \right)^{-1},
\end{equation}
which at worst overestimates $\chi(x)$ by 24\% (when $x \approx 1$), well within the required accuracy for most estimates. However, for an even better approximation, one can instead use
\begin{equation}
	\label{eq:ChiApproxBetter}
	\chi(x) \approx \left( 1 + \left(\frac{x}{\sqrt{2}}\right)^{\sqrt{2}} \right)^{-\sqrt{2}},
\end{equation}
which at worst underestimates $\chi(x)$ by only about 8\% (also when $x \approx 1$).

Similarly, we can combine the expressions for long and short bunches (Eqs.~\ref{eq:LongBunchMaxField} and \ref{eq:ShortBunchMaxField}). These are related by
\begin{equation}
	g_{\max}^{\mathrm{long}} = \frac{1}{\sqrt{8\pi} k_p^2 \sigma_z^2} g_{\max}^{\mathrm{short}},
\end{equation}
which means that the smallest of the two expressions can be approximated by
\begin{equation}
	\min\left(g_{\max}^{\mathrm{long}}, g_{\max}^{\mathrm{short}}\right) \approx \left(1 + \sqrt{8\pi} k_p^2 \sigma_z^2 \right)^{-1} g_{\max}^{\mathrm{short}}.
\end{equation}

Lastly, since the field gradient from the blowout regime represents a saturation, it is sensible to keep the $\min$-function in the final simplified expression for the maximum plasma wakefield-based focusing gradient  
\begin{equation}
	\label{eq:MaxFieldSimplified}
	g_{\max} \approx -\frac{e \mu_0 c}{2} \min\left( n_0, \frac{N k_p^2 \sigma_z}{\pi \sigma_r^2 \left(1+\frac{k_p^2 \sigma_r^2}{2} \right) \left(1 + \sqrt{8\pi} k_p^2 \sigma_z^2 \right)}\right),
\end{equation}
where we have used the identity $c^2 = 1/\mu_0\epsilon_0$ for Eq.~\ref{eq:IntenseBunchMaxField}. Keep in mind that this can be be made more accurate by substituting Eq.~\ref{eq:ChiApproxBetter} for Eq.~\ref{eq:ChiApprox}.

\begin{figure*}[ht]
	\centering\includegraphics[width=0.97\textwidth]{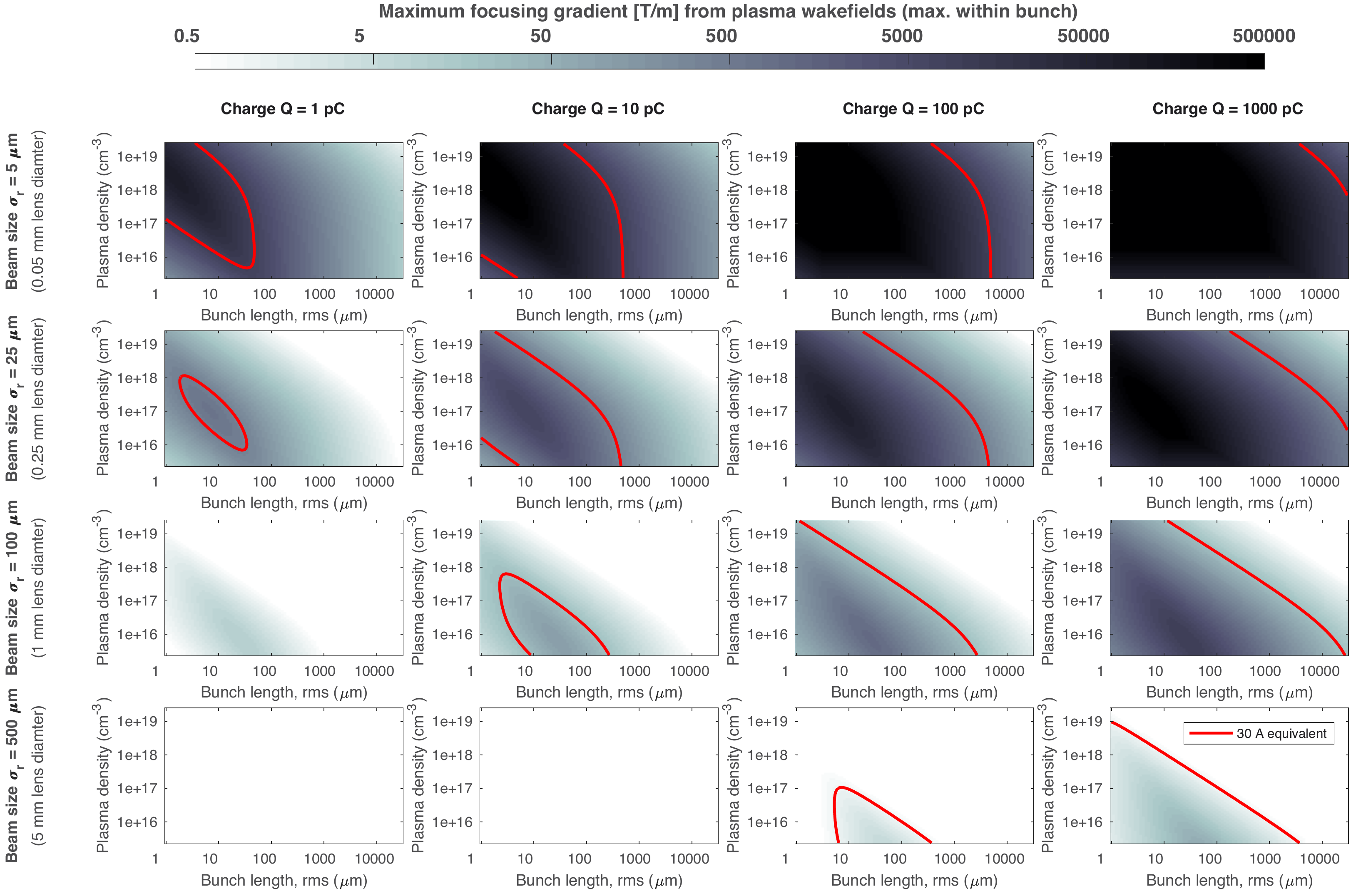}
	\caption{Visualization of the entire relevant 4D parameter space. The 2D grid of density plots shows varying beam size (outer vertical) and beam charge (outer horizontal), and within each plot the maximum focusing gradient within the bunch due to plasma wakefields is shown against the plasma density (inner vertical) and bunch length (inner horizontal). The red line indicates the contour equivalent to a 30~A current active plasma lens (3\% of a 1~kA peak current), which can be used as an approximate upper bound for the allowed plasma wakefields. Note that for the calculation of the gradient of the active plasma lens, the radius of the capillary is constrained to be 10 times the rms beam size. It is apparent that the combination of high charge and either small beam size or short bunch length largely excludes the use of a distortion-free active plasma lens.}
    \label{fig:Limits4D}
\end{figure*}

\subsection{Discharge current and beam size lower limits}
The requirement for operating an active plasma lens without deleterious effects from plasma wakefields can be formulated as
\begin{equation}
	g_{\mathrm{APL}} \gg g_{\max}.
\end{equation}
Although there is not much to be gained from expanding and simplifying this inequality, it can be recast as a minimum requirement for the discharge current in the active plasma lens
\begin{equation}
	I \gg e c \min\left( n_0 \pi R^2, \frac{N k_p^2 \sigma_z n_R^2}{\left(1+\frac{k_p^2 R^2}{2 n_R^2} \right) \left(1 + \sqrt{8\pi} k_p^2 \sigma_z^2 \right)} \right),
\end{equation}
where the right hand side represents the equivalent current needed to drive an active plasma lens of the same gradient as that of the beam-driven plasma wakefields. Here we have fixed the number of beam size sigmas $n_R = R/\sigma_r$ desired inside the capillary radius, as it is desirable to use the smallest possible capillary to maximize the current density. Note, however, that there are manufacturing limitations to the minimum capillary diameter, as well as a limit to the current density due to nonlinearities \cite{vanTilborgPRAB2017}.

Alternatively, the above expression can be recast as the minimum required beam size to avoid plasma wakefield distortions
\begin{equation}
	\label{eq:MinBeamSize}
	\sigma_r \gg \sqrt{\frac{2 c Q n_R^2}{I\left(\frac{1}{\sigma_z} + \sqrt{8\pi} k_p^2 \sigma_z \right)} - \frac{2}{k_p^2}}.
\end{equation}
In this case, the beam will almost always be large enough to not create a nonlinear blowout, and we can therefore make use of only the latter argument in the min-function (the linear regime).

\section{Parameter space visualization}

As can be seen from Eq.~\ref{eq:MaxFieldSimplified}, the maximum plasma wakefield focusing gradient depends only on four parameters: three beam parameters (bunch length, charge and beam size) and one plasma parameter (plasma density). Each of these parameters have practical limitations to their span. Bunch lengths typically range from 1~$\mu$m (laser wakefield injection) to about 1~m (microtrons or proton beams), and bunch charge typically varies between 1~pC and 1~nC. The beam size is more flexible than the latter two, but will typically be around 1~$\mu$m to around 1~mm, depending on the application. Lastly, the plasma density in an active plasma lens is limited by the Paschen Law \cite{PaschenADP1889}, which for a lens of cm-to-m scale will only allow breakdowns in the range 0.1-1000~mbar or equivalently around $2.4 \times 10^{15}$~cm$^{-3}$ to $2.4 \times 10^{19}$~cm$^{-3}$.

Because of Eq.~\ref{eq:MaxFieldSimplified}, this 4D parameter space can finally be visualized in its entirety in a 2D grid of density plots, as shown in Figure~\ref{fig:Limits4D}. In general, the maximum focusing gradient is stronger for higher charge beams and for smaller beam sizes. Plasma density, bunch length and beam size are all interdependent: the strongest focusing occurs when the bunch length matches the plasma wavelength, resulting in a peak whose position in ($n_0$,~$\sigma_z$)-space depends on the beam size. We observe that plasma wakefields presents a significant challenge to active plasma lenses if used for intense bunches, whether due to a short bunch length, high charge or small beam size.

\section{Applications}

The ability to quickly calculate and visualize the full parameter space allows us to estimate whether active plasma lenses are suitable for various applications, and if so, with what constraints. Perhaps the most important such application is that of the linear collider. Before we investigate that, however, it is valuable to compare the analytic expression to recent active plasma lens experiments.

\begin{table*}[ht]
	\centering
    \begin{tabular}{l l l l l l}
    \hline
     	\textbf{Experiment} & \textbf{LBNL} & \textbf{INFN} & \textbf{DESY}  & \textbf{INFN}& \textbf{CERN}  \\
     	& BELLA & Frascati \#1 & Mainz & Frascati \#2 & CLEAR  \\
    \hline
        Energy (MeV) & 62 & 126 & 855 & 127 & 200 \\
        Charge (pC) & 10-50 & 50 & 1 & 50 & 1-1000 \\
        Bunch length ($\mu$m) & 2 & 330 & $\sim$10$^6$ & 350 & 300-1500 \\
        Beam size, rms ($\mu$m) & 600-2100 & 130 & 150 & 95 & 30-70 \\
        Capillary radius ($\mu$m) & 500 & 500 & 500 & 500 & 500 \\
        Capillary length (mm) & 15 & 30 & 7-33 & 30 & 15 \\
        Peak current (A) & 440 & 93 & 740 & 95 & 450-750 \\
        Gas pressure (mbar) & 3.3 (He) & 40 (H$_2$) & 4 (H$_2$) & 300 (H$_2$) & 4-50 (Ar) \\
        Plasma density (cm$^{-3}$) & 8$\times$10$^{16}$ & 9$\times$10$^{16}$ & 10$^{17}$ &  $\leq 6$$\times$10$^{16}$ & $\leq 10^{18}$ \\
    \hline
        Active plasma lens gradient (T/m) & 350 & 74 & 590  & 76& 360-600 \\
        Max.~gradient from wakefields (T/m) & 10$^{-4}$-0.01 & 3.5 & 4$\times$10$^{-5}$ & $\lesssim$ 180 & $\lesssim$ 7000 \\
    \hline
    \end{tabular}
    \caption{Beam and plasma lens parameters for recent and current active plasma lens experiments. Also shown is the predicted active plasma lens focusing gradient (using Eq.~\ref{eq:TransGradient}) as well as the maximum focusing gradient within the bunch caused by plasma wakefields (using Eq.~\ref{eq:MaxFieldSimplified}). Three of the experiments (LBNL \cite{vanTilborgPRAB2017}, INFN Frascati \#1 \cite{PompiliAPL2017} and DESY \cite{RockemannPrivComm}) have reported no plasma wakefield distortion, consistent with expectation. However, a dedicated passive plasma lensing experiment (INFN Frascati \#2 \cite{MarocchinoAPL2017}) has observed the effect, which should be even stronger in the CERN-experiment \cite{LindstromNIMA2018}. In these latter experiments, the subsequent plasma decay after the current discharge is used to scan the plasma density, which couples most strongly to the $\sim$300 $\mu$m long bunches at around 10$^{14}$ cm$^{-3}$ (used to calculate of the wakefield gradient).}
    \label{tbl:Experiments}
\end{table*}

\subsection{Recent active plasma lens experiments}
In recent years, several experiments (by LBNL \cite{vanTilborgPRAB2017}, INFN \cite{PompiliAPL2017,MarocchinoAPL2017} and DESY \cite{RockemannPrivComm}) have built and tested active plasma lenses. Table~\ref{tbl:Experiments} shows the experimental parameters for each of these experiments, and predicts the maximum wakefield focusing gradient within the bunch. Comparing this gradient to the expected active plasma lens gradient, we find that many of the experiments should have seen no significant plasma wakefield focusing, consistent with their observations. However, a dedicated passive plasma lensing experiment at INFN Frascati \cite{MarocchinoAPL2017} did observe wakefield focusing, again in accordance with expectation. An ongoing experiment \cite{LindstromNIMA2018} in the CLEAR User Facility \cite{GambaNIMA2018} at CERN will further probe the limits set by these plasma wakefields, over a wide range of bunch and plasma parameters.

\subsection{Future colliders}

\begin{figure*}[ht]
	\centering\includegraphics[width=\textwidth]{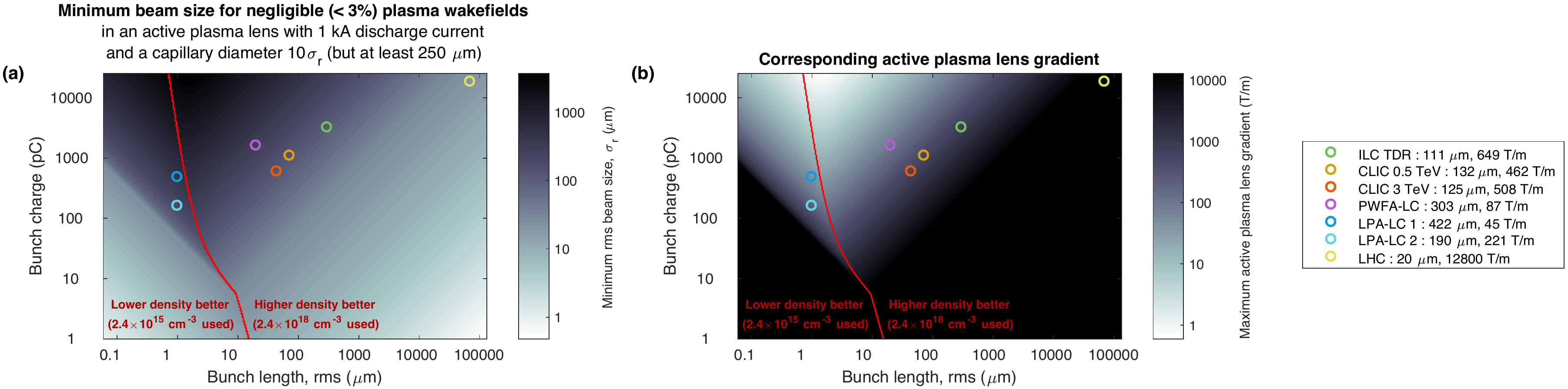}
	\caption{(a) Minimum beam size required in an active plasma lens to have negligible distortion, i.e.~when the gradient from plasma wakefields is 3\% or less of the active plasma lens gradient, and (b) the corresponding active plasma lens gradient given this size. Note that the capillary diameter is constrained to 10~times the beam size, such that a smaller beam size gives a larger gradient (Eq.~\ref{eq:TransGradient}), but that this lens diameter is constrained to be at least 250~$\mu$m. The parameter space is divided in two parts, where lower (left) and higher (right) plasma densities allow smaller beam sizes, respectively. A typical discharge current of 1~kA is used, but smaller beam sizes can be tolerated if this current is increased. Collider parameters from Table~\ref{tbl:Colliders} are indicated as colored circles.}
    \label{fig:MinBeamSize}
\end{figure*}

\begin{table*}[ht]
	\centering
    \begin{tabular}{l l l l l l l l}
    \hline
     	\textbf{Collider} & \textbf{ILC} & \textbf{CLIC} & \textbf{CLIC} & \textbf{PWFA-LC} & \textbf{LP-LC} & \textbf{LP-LC} & \textbf{LHC} \\
        & TDR & 0.5~TeV & 3~TeV & 1~TeV & Ex.~1 & Ex.~2 & 13~TeV \\
    \hline
        Final beam energy (GeV) & 250 & 250 & 1500 & 500 & 500 & 500 & 6500 \\
        Charge per bunch (pC) & 3200 & 1088 & 595 & 1600 & 480 & 160 & 18400 \\
        Bunch length, rms ($\mu$m) & 300 & 72 & 44 & 20 & 1 & 1 & 7.6$\times$10$^{4}$ \\
        Normalized emittance, x/y ($\mu$m rad) & 10/0.035 & 2.4/0.025 & 0.66/0.02 & 10/0.035 & 1/0.01 & 1/0.01 & 3.75 \\
    \hline
        \multicolumn{8}{l}{\textit{Considerations for an active plasma lens with 1~kA discharge current and a minimum diameter 250~$\mu$m}}\\
        Min.~beam size for negligible ($<$ 3\%) wake ($\mu$m) & 111 & 132 & 125 & 303 & 422 & 190 & 20 \\
        Max.~APL gradient with negligible wake (T/m) & 649 & 462 & 508 & 87 & 45 & 221 & 12800 \\
        Required beta function $\sqrt{\beta_x \beta_y}$, final energy (m) & 1.0$\times$10$^{4}$ & 3.5$\times$10$^{4}$ & 4.0$\times$10$^{5}$ & 1.5$\times$10$^{5}$ & 1.7$\times$10$^{6}$ & 3.5$\times$10$^{5}$ & 0.74 \\
    \hline
    \end{tabular}
    \caption{Beam parameters for proposed linear colliders, including the existing circular collider LHC for reference. Constraints on the minimum beam size and beta function in an active plasma lens and the maximum gradient it can provide are shown, based on Fig.~\ref{fig:MinBeamSize} (assuming 1~kA discharge current and minimum diameter 250~$\mu$m). For ILC and CLIC the beta function is too large to allow use of plasma lenses in the main linac, and even more so in the plasma-based collider concepts (PWFA-LC and LP-LC). However, these large beta functions are compatible with use as an alternative for the final doublet in a final focus system.}
    \label{tbl:Colliders}
\end{table*}

One of the main motivations for the recent surge of interest in active plasma lenses is to use them for compact staging \cite{SteinkeNature2016} of plasma accelerators, an important step towards a plasma-based ultra-compact linear collider. However, to maximize collisions at the interaction point, collider beams typically have both high charge and a short bunch length. If active plasma lenses are to be used in a linear collider without suffering distortion from plasma wakefields, the beam size needs to be sufficiently increased. 

Table~\ref{tbl:Colliders} lists the beam parameters for various proposed linear colliders, both mature designs (ILC \cite{ILCTDR2013} and CLIC \cite{CLICCDR2012}) and early concepts for plasma-based colliders (PWFA-LC: a beam-driven plasma wakefield linear collider \cite{AdliCSS2013}, and LP-LC: a laser-plasma linear collider \cite{Schroeder2008}), as well as the existing circular collider LHC \cite{LHCReport2014}. For each machine, a minimum beam size can be estimated by requiring e.g.~3\% or less distortion from plasma wakefields (i.e.~extra focusing in addition to that from the current). 

Clearly, use of smaller beam sizes can be compensated by use of a higher discharge current, which is limited in two ways. Given a constant total current, the current density and hence focusing gradient will increase with decreasing capillary diameter and vice versa. However, if the current density is too large, radial nonuniformities form in the active plasma lens itself through plasma heating and pinching effects \cite{vanTilborgPRAB2017,BoggashPRL1991}. Therefore, an alternative is to instead consider constant current density. However, this is again unrealistic for large capillary diameters, where at some point the total current is beyond what can be supplied. Since it is not yet clear what these limits are quantitatively, we will consider as an example the combination of a typical current of 1~kA and a minimum lens diameter of 250~$\mu$m, based on the smallest capillaries used for active plasma lensing \cite{vanTilborgPRL2015}.

Figure~\ref{fig:MinBeamSize} shows both the minimum beam size and corresponding active plasma lens gradient for the parameter space relevant to linear colliders. For each combination of bunch length and charge, we are free to choose a plasma density to minimize plasma wakefields. This divides the parameter space in two parts: one which favors lower densities and one which favors higher densities, where the equivalent of 0.1 mbar and 100 mbar gas pressures were used, respectively. The collider parameters from Table~\ref{tbl:Colliders} are indicated in the plots, and are seen to mostly require beam sizes of approximately 100-400~$\mu$m rms. 

We observe that of all the linear collider designs listed in the example, the ILC is most promising with close to a kT/m of focusing gradient. On the opposite end, both the plasma wakefield accelerator-based collider designs require large capillaries (2-4~mm diameter) reaching only about 100~T/m. However, while large capillaries and weak focusing is not ideal, these machines also require low emittances, which means that a large beam size is equivalent to a very large beta function of order 10$^{4}$-10$^{6}$ m (see Table~\ref{tbl:Colliders}). Using an active plasma lens is therefore not practical in the main linac, where a significantly smaller beta function (1-100~m) is required: in ILC and CLIC this is needed to reduce alignment tolerances, whereas in plasma-based colliders it is to avoid large chromaticity in the staging optics \cite{LindstromNIMA2016} between plasma cells. However, if discharge currents can be increased to 10-100~kA and emittance growth from pinching effects can be avoided, use of active plasma lenses may in fact be a viable option for main linac transport.

On the other hand, large beta functions of this order occurs naturally in the final focusing system \cite{RaimondiPRL2001}, which may make a large diameter plasma lens a suitable replacement for the final quadrupole doublet. Note that it is not mainly the increase in focusing gradient which represents the improvement in this case, but instead the ability to focus in both planes simultaneously. This would remove the need to defocus one plane enormously before the interaction point, which could greatly reduce the chromaticity and potentially allow the final focus system to be shortened.

Finally, while usage in currently proposed linear colliders may prove challenging, Fig.~\ref{fig:MinBeamSize} indicates that active plasma lensing can be suitable for machines with a different temporal structure and reduced intensity. This may include e.g.~proton bunches like those produced by the SPS and LHC, with higher bunch charge, but significantly longer bunch lengths, in irradiation or proton therapy \cite{WilsonRad1946,NewhauserPMB2015} facilities, or even radically different advanced collider concepts like that based on the dielectric laser accelerator \cite{EnglandRMP2014}, with ultra-low bunch charge, but a very high (MHz) bunch repetition rate.

\section{Conclusion}
In this paper, we developed analytic expressions for the maximum focusing gradient within a bunch due to plasma wakefields (Eq.~\ref{eq:MaxFieldSimplified}). Verified by PIC simulations and consistent with experiments, these expressions allow fast exploration of the full parameter space, which indicate that active plasma lenses are mainly suitable for low charge or long beams, with an exception for extremely ultra-short beams (bunch lengths less than $\sim$1 $\mu$m). This means that for currently proposed parameters, active plasma lenses are not suitable for use in a linear collider main linac, nor in staging of plasma wakefield accelerator cells, unless discharge currents much greater than 1~kA are used. Fortunately, there are still possibilities for these lenses to replace the final doublet in a conventional final focusing system, to be used in proton machines or even in novel collider designs with an alternative timing structure. Nevertheless, the analytic expressions developed provide fundamental constraints for any such consideration, and should be used as a guide in future accelerator designs involving active plasma lenses.

\section{Acknowledgements}
The authors wish to thank Kyrre N.~Sj{\o}b{\ae}k, Jens Osterhoff, Spencer Gessner and S\'ebastien Corde, as well as the CLIC Novel Accelerator Working Group at CERN for useful discussions. This work was supported by the Research Council of Norway (Grant No.~230450).


\end{document}